\documentclass[12pt]{iopart}
\usepackage{amssymb,amsfonts}
\usepackage{latexsym}

\renewcommand\[{\begin{equation}}
\renewcommand\]{\end{equation}}
\newcommand{\al}{\alpha}
\newcommand{\bt}{\beta}
\newcommand{\ga}{\gamma}
\newcommand{\da}{\delta}
\newcommand{\n}{\nabla}
\newcommand{\ba}{\begin{eqnarray}}
\newcommand{\ea}{\end{eqnarray}}

\newcommand{\LT}{\left[}
\newcommand{\RT}{\right]}

\newcommand{\cO}{{\cal O}}
\newcommand{\cF}{{\cal F}}
\newcommand{\Ra}{\Rightarrow}

\begin{document}
\title[Ghost-free gravity]{Generalized ghost-free quadratic curvature gravity}


\author{Tirthabir Biswas$^1$, Aindri\'u Conroy$^2$, Alexey S. Koshelev$^3$ and
Anupam Mazumdar$^2$}


\address{$^1$ Department of Physics, Loyola University, New Orleans, LA 56302}
\address{$^2$ Consortium for Fundamental Physics, Lancaster University,
Lancaster, LA1 4YB, UK}
\address{$^3$ Theoretische Natuurkunde, Vrije Universiteit Brussel and The
International Solvay Institutes, Pleinlaan 2, B-1050, Brussels, Belgium}

\ead{tbiswas@loyno.edu,
a.conroy@lancaster.ac.uk,
alexey.koshelev@vub.ac.be,
a.mazumdar@lancaster.ac.uk}

\begin{abstract} In this paper we study the most general covariant action of
gravity
up to terms that are quadratic in curvature. In particular this includes
non-local, infinite
derivative theories of gravity which are {\it ghost-free} and exhibit {\it asymptotic freedom} in the ultraviolet.
We provide a detailed algorithm for
deriving the equations of motion for such actions containing an arbitrary number of the covariant D'Alembertian operators, and this is our main result. We also perform a number of tests on the field equations we derive,
 including checking the Bianchi identities and the weak-field limit. Lastly, we
consider the special subclass of ghost and asymptotically free theories of
gravity by way of an example.
 \end{abstract}


\submitto{\CQG}

\maketitle

%
\section{Introduction}
\label{sec:intro}

Einstein's general relativity (GR) is a remarkable theory in many respects but
there still remain some major unresolved questions. Despite a vast amount of observational data in favour of GR \cite{will}, the presence of cosmological and
black hole singularities are examples of problems at the classical level which strongly suggest the incompleteness (or inconsistency) of GR in the ultraviolet (UV).
At the quantum level GR is not UV complete. Modification of GR is demanded in this regard
but one must take care of preserving the conformance with the available
data. Indeed, the above
mentioned classical and quantum problems could be closely related as both concern short-distance physics.

Most of the work on modifying GR has concentrated upon studying finite higher order (derivative) gravity such as ``Fourth Order Gravity'' which are quadratic in curvatures:
$$ {\cal L}\sim R+f_1R^2+f_2R_{\mu\nu}R^{\mu\nu} +f_3R^{\mu\nu\lambda\sigma}R_{\mu\nu\lambda\sigma},$$
where $\mu,~\nu=0,1,2,3$ and $f_1,f_2,f_3$ are appropriate constant coefficients - a particular variant of which give the famous
Gauss-Bonnet gravity~\cite{lovelock}.

The above action naturally yields $4$ derivatives and it was studied
extensively
by Stelle in Refs.~\cite{Stelle:1976gc,Stelle:1977ry,Quandt:1990gc}. One of the most interesting properties of $4$th derivative gravity is that it can be made renormalizable~\cite{Stelle:1976gc},
thus in the UV the modified graviton propagator leads to convergence of the Feynman diagrams at 1-loop and
beyond. However, this comes at the cost of the presence of the ``Weyl'' ghost in the tensor component of the modified graviton propagator~\cite{Stelle:1976gc}.

The authors Biswas, Mazumdar and Siegel (BMS) in Ref.~\cite{BMS}  argued that
the absence of ghosts in the modified propagator coupled with asymptotic freedom
can {\it only} be realized if one considers an infinite set of higher derivative
terms, in particular in the form of an exponential, that are allowed by general
covariance, see also~\cite{nlgauge}. Although, the infinite derivative action
considered in~\cite{BMS}  was only partially successful in realizing asymptotic
freedom, it was able to find cosmological non-singular (without the Big
Crunch/Bang singularity) bouncing solutions in these type of theories. Based on
the action of BMS, cosmological perturbation analyses were performed in
Refs.~\cite{BKM} which further demonstrate the robustness of the bouncing model and its possible connection to inflationary cosmology\footnote{For
cosmological applications of infinite derivative scalar field theories, see for
instance~\cite{cosmo-non-local} and references therein.}. All this strongly
suggests
that it may be possible to weaken  gravitational interactions in a consistent manner at short distances and at early times.

Indeed, recently it was shown by Biswas, Gerwick, Koivisto and Mazumdar (BGKM) in Ref.~\cite{Biswas:2011ar} that gravity in the UV can be made
asymptotically free without violating basic principles of physics, such as
unitarity and general covariance by including appropriate infinite set of higher
derivative terms. The
action considered in Ref.~\cite{Biswas:2011ar} is of the form
\begin{equation}\label{action}
\fl S=\int d^{4}x\sqrt{-g}\left(\frac{R}{2}+R{\cal
F}_{1}(\Box)R+R^{\mu\nu}{\cal
F}_{2}(\Box)R_{\mu\nu}+C^{\mu\nu\lambda\sigma}{\cal
F}_{3}(\Box)C_{\mu\nu\lambda\sigma}\right)\,,
\end{equation}
where ${\cal F}_{i}(\Box)$ are functions of the D'Alembertian
operator, $\Box=g^{\mu\nu}\nabla_{\mu}\nabla_{\nu}$, $\nabla_{\nu}$  is the
covariant derivative and $g_{\mu\nu}$ is $4$ dimensional metric with
$(-,+,+,+)$, $R_{\mu\nu}$ and $C_{\mu\nu\lambda\sigma}$ are the Ricci and Weyl
tensors~\footnote{In the original
action introduced in Ref.~\cite{Biswas:2011ar}, we
had the Riemann tensor rather than the Weyl tensor. The Weyl tensor
however is
identically zero on conformally flat manifolds and therefore it is a more
convenient choice
for most calculations.} respectively which are given in terms of the Riemann
tensor, $R_{\mu\nu\lambda\sigma}$, as usual via
\[
\label{weyl}
C_{\;\alpha\nu\beta}^{\mu}=R_{\;\alpha\nu\beta}^{\mu}-\frac{1}{2}(\delta_{\nu}^{
\mu}R_{\alpha\beta}-\delta_{\beta}^{\mu}R_{\alpha\nu}+R_{\nu}^{\mu}g_{
\alpha\beta}-R_{\beta}^{\mu}g_{\alpha\nu})+\frac{R}{6}(\delta_{\nu}^{\mu}g_{
\alpha\beta}-\delta_{\beta}^{\mu}g_{\alpha\nu})
\]
We have also set the reduced Planck mass to one, and  we only consider ${\cal
F}_{i}(x)$'s that are analytic at $x=0$, otherwise one does not recover GR in
the infrared, see~\cite{Biswas:2011ar} for details.
Further analytic stable cosmological bouncing solutions for action
~(\ref{action}) were
found in \cite{Koshelev:2013lfm}.

Gravity being a gauge theory, i.e. diffeomorphism invariant, should allow all
possible gauge invariant terms in its ``effective action'' and (\ref{action}),
at least, contains all possible terms with at most two curvatures. We emphasize
that ${\cal F}_{i}(x)$'s can be  transcendental functions containing an infinite
set of derivatives and this is of course the case that is of most interest to
us. We should point out that there has been a growing interest in recent years
in such/similar infinite derivative gravitational actions deriving its
motivation from cosmology~\cite{cosmology}, renormalization group
flows~\cite{RGflow} and quantum gravity~\cite{QG}. We also find it curious to
note that such infinite higher derivative actions appear
in non-perturbative string theories, see for instance Ref.~\cite{sft}.
 The higher derivative corrections essentially take into account the $\alpha'$
corrections in string theory. Unfortunately, however, close string theories
containing graviton physics do not yet provide an action with all orders in
$\alpha'$.

Let us now briefly expound upon the advantages of having an infinite set of
derivatives.  In GR the graviton propagator becomes
\begin{equation}\label{prop-1}
{\Pi}_{{GR}}\sim \frac{P_2}{k^2}-\frac{P_s}{2 k^2}\,.
\end{equation}
where $P_2$ and $P_s$ refers to the projection operators corresponding to the
massless spin-2 and the scalar degree of freedom in the metric, see
Ref.~\cite{VanNieuwenhuizen:1973fi} for details. The graviton propagator in BGKM
extension (\ref{action}) of gravity modifies
to~\cite{Biswas:2011ar,Biswas:2013ds} 
\begin{equation}\label{prop-2}
{\Pi}_{BGKM}\sim
\frac{1}{f(k^2)}\left(\frac{P_2}{k^2}-\frac{P_s}{2k^2}\right)\,,~\mathrm{where}
~f(k^2)\rightarrow 1~\mathrm{when}~k^2\rightarrow 0\,.
\end{equation}
In order for the propagator to be ghost-free, $f(k^2)$ must have no
zeros on the complex plane\footnote{Such functions can be represented as the
exponent of an entire function. For a polynomial
$f(k^2)$ extra poles appear, an extra degree of freedom other than the massless
graviton, and one generates new degrees
of freedom also in the equation of motion, which due to Ostrogradski analysis
will be ghost-like in nature~\cite{ostrogradski}.
For higher derivative theories the initial condition problem has also been
addressed in Ref.~\cite{barnaby}.}, so that
there are no extra poles in the propagator.
However, at low energies, $k^2\rightarrow 0$, the above equation~(\ref{prop-2})
asymptotes to the GR
propagator, i.e. ~(\ref{prop-1}). An example of $f(k^2)$ which leads to
a ghost and asymptotically free theory is given
by~\cite{Biswas:2011ar,Biswas:2013ds}
\begin{equation}
{f}(k^2)=e^{\gamma k^2}\,,
\end{equation}
with $\gamma>0$. The propagator is exponentially suppressed in the UV leading to
softening of the gravitational force at short distance.

So far, based on linearized equations~\cite{Biswas:2011ar}, we have found that
non-local asymptotically free actions can indeed resolve black hole
singularities, although the argument only holds for mini black holes with a
mass much below the Planck mass~\footnote{Similar argument holds for linearized
gravitational waves in GR, where the aptitude of the gravitational wave diverges
near the source, but
with BGKM propagator the amplitude of the gravitational waves remain finite near
the source throughout the domain while recovering that of the GR's prediction in
the IR~\cite{Biswas:2011ar}.}. Further, the theory also admits  cyclic
cosmologies with nonsingular transitions from expansions to contractions
(turnarounds) and vice versa (bounces). However, to study the robustness of the
resolutions of these classical singularities one needs to go beyond the
linearized  equations of motion.  This is the primary motivation for the present
work where we obtain the field equations for the most general higher derivative
action containing at most two curvatures, and generalize the results presented
in ~\cite{BMS,BKM,Koshelev:2013lfm}. A topical  application of this would be to
improve on
Starobinsky's model of inflation~\cite{Starobinsky:1980te}, for a review on
primordial inflation see~\cite{Mazumdar:2010sa}, where only
$${\cal L} \sim R +R^2$$ terms were considered, but clearly there is no reason
why other quadratic curvature terms would be missing! For example, no underlying
symmetry is known which  could prevent these terms from appearing in the
``effective action'', and therefore one should really consider an action of the
form ~(\ref{action}). Actually, including such terms can lead to interesting
phenomenological consequences:  since such modifications are expected to produce
bounce with a solution $ a(t) =a_0\cosh(\kappa t)$, where $a_0$ and $\kappa$ are
positive constants~\cite{BMS,BKM,Koshelev:2013lfm}, which inevitably entail a
super-inflationary phase near the bounce, such pre-inflationary dynamics may be
able to account for the low multipoles that has been observed in
CMB~\cite{Biswas:2013dry}.

For most part, our paper is technical and we provide a detailed account of the
methodology we used in taming actions such as (\ref{action}). In fact, our
algorithm for dealing with higher derivative actions involving an arbitrary
number of $\Box$ operators can be applied more generally to any action of the
form
$$S=\int d^4x\ S{\cal F}(\Box) T$$
where $S$ and $T$ are any tensors constructed out of the Riemann curvatures and
the metric. We note in passing that over the last decade or so higher derivative
gravitational actions have gained a lot of attention as alternatives to the dark
energy paradigm (for a review see \cite{Deser:2013uya}), and some of the general
techniques/results in our paper may be relevant for such investigations.

Our paper is organized as follows: Section~2 is devoted to deriving the field
equations for ~(\ref{action}). Section~3 deals with various checks of our
results. Firstly, we
verify that our results reproduce the field equations obtained in previous
literature for some special cases. Secondly, we
check that
the Bianchi identities are satisfied. Although a somewhat arduous task,
this is a strong indicator of the validity of our derivation; and finally we
reduce our result in the weak-field limit by looking at
perturbations around Minkowski space-time and analysing the Newtonian potentials
and comparing with known expressions. In Section~4, we consider a particular
subclass of quadratic curvature theories of gravity that are both ghost and
asymptotically free, before concluding by summarizing our results and providing
a brief outlook of possible future research. Finally, in the appendices, we give
explicit details of the derivations of the equation of motion (Appendix A), the
details of the Bianchi identity calculation (Appendix B) and the equation of
motion in a form useful for comparison with sixth order gravity \cite{Decanini}
(Appendix C).
\section{Action and equations of Motion}
Let us start by recalling that the most general generally-covariant gravity
action with at most  quadratic curvature terms  is of the
form~\cite{Biswas:2011ar}
\[
S_{q}=\int d^{4}x\sqrt{-g}R_{\mu_{1}\nu_{1}\lambda_{1}\sigma_{1}}{\cal
O}_{\mu_{2}\nu_{2}\lambda_{2}\sigma_{2}}^{\mu_{1}\nu_{1}\lambda_{1}\sigma_{1}}R^
{\mu_{2}\nu_{2}\lambda_{2}\sigma_{2}}\ .
\]
 As discussed in~\cite{Biswas:2011ar}
we
need only the graviton propagator to understand both the asymptotic behaviour
of gravity in the UV and the potential problem with ghosts. The focus
of~\cite{Biswas:2011ar} was
on linearized solutions of gravity around the Minkowski background
\[
g_{\mu\nu}=\eta_{\mu\nu}+h_{\mu\nu}\,,
\]
and accordingly one only needed to consider an action that
contained  terms which are quadratic in $h_{\mu\nu}$.

We now want to expand our treatment to include fluctuations around an arbitrary
background for the most general quadratic curvature gravity action given by
~(\ref{action}).   For simplicity we
often omit hereafter the explicit dependence of ${\cal F}_i(\Box)$ on $\Box$
writing simply ${\cal F}_i$.

\subsection{General Methodology}
\subsubsection{Single $\Box$}
In order to elucidate the methods involved in finding the field equations, we
will start with a simple example:
\[
S_{p}=\int d^4x\sqrt{-g} T\Box S\,,
\]
where $S$ and $T$ are arbitrary scalars made out of the Riemann and the metric
tensors. Varying this gives us
\[
\delta S_{p}=\int d^4x\ \sqrt{-g}\left({h\over 2} T\Box S+ \delta T\Box
S+T\delta(\Box) S+T\Box \delta S\right).
\]
where  we have defined
\[
\delta g_{\mu\nu}=h_{\mu\nu}\,,
\]
so that~\footnote{The indices are always raised and lowered with respect to the
background metric $g_{\mu\nu}$.}
\[
\delta g^{\mu\nu}=-h^{\mu\nu}\,.
\]
Variations of the various curvatures, such as  $\delta R$, $\delta R_{\mu\nu}$
and $\delta
C_{\mu\nu\lambda\sigma}$ are well known and are given in
Appendix~A for completeness.
Any composite function of these can be straightforwardly computed with the help
of the individual variations.
Calculation of $\delta(\Box)S$, however, is a little
trickier and we will provide it below. The derivation of
$\delta(\Box)S_{\mu\nu}$ and $\delta(\Box)S_{\mu\nu\lambda\sigma}$, which we
will also need, can be found
in  Appendix~A as well.

From the definition of the D'Alembertian operator
 we have
\[
\delta(\Box)S=-h_{\alpha\beta}S^{;\alpha;\beta}+g^{\mu\nu}\delta(\nabla_{\mu})S_
{;\nu}+g^{\mu\nu}[\delta(\nabla_{\nu})S]_{;\mu}\,,
\label{deltaBoxS}
\]
where semicolons denote the covariant derivative.
Next, by using the general definition of the covariant derivative of a tensor
and treating $S_{;\alpha}$ as a $(0,1)$-tensor one finds that
\[
g^{\mu\nu}\delta(\nabla_{\mu})S_{;\nu}=-g^{\mu\nu}\delta\Gamma_{\mu\nu}^{\lambda
}S_{;\lambda}\,.
\]
The trick is to observe that the $\delta$-variation operates only on the
Christoffel symbols and not upon $S$ and in this case the final term vanishes as
$S$ and $T$ are scalars. We then arrive at an important result:
\[
\label{deltaR}
\delta(\Box)S=-h_{\alpha\beta}S^{;\alpha;\beta}+\frac{1}{2}g^{\alpha\beta}S_{
;\lambda}(h_{\alpha\beta})^{;\lambda}-S^{;\alpha}(h_{\alpha\beta})^{;\beta}.
\]
The variation of the $S_p$ can now be written in a convenient form
\[
\delta S_{p}=\int d^4x\ \sqrt{-g}\left(g^{\mu\nu} T\Box S+ \Box S{\delta T\over
\da g_{\mu\nu}}+ \Box T{\delta S\over \da g_{\mu\nu}} \right) h_{\mu\nu}\,,
\]
where we have liberally used integration by parts to transfer the covariant
derivatives acting on $h_{\mu\nu}$ to $S,T$.
\subsubsection{Multiple $\Box$'s}
What is crucial, of course, is that the above manipulations can be straight
forwardly generalized to the case when one has many $\Box$ operators.
In particular, by repeated integration by parts one can show that the result in
equation
(\ref{deltaR}) is useful for all powers of the D'Alembertian operator and indeed
any function ${\cal F}_i(\Box)$. We have
\[
\int d^4x\ \sqrt{-g} T\delta(\Box^{n})S=\sum_{m=0}^{n-1}\int d^4x\ \sqrt{-g}
\square^{m}T\delta(\square)\square^{n-m-1}S\,,
\]
which is itself simply a result of repeated integration by parts. Its equivalent
in terms of an arbitrary function ${\cal F}_i(\Box)$ is as follows
\ba
\label{deltaboxn}
\int d^4x\sqrt{-g}T\delta
F_{i}(\Box)S=\sum_{
n=1}^{\infty}\sum_{m=0}^{n-1}\int
d^4x\sqrt{-g}f_{i_{n}}\square^{m}T\delta(\square)\square^{n-m-1}S\,,
\ea
where $f_{i_n}$'s are the Taylor series expansion
coefficients:
\[
\label{F_i}
{\cal F}_i(\Box)=\sum_{n=0}^{\infty}f_{i_n}\Box^n\ .
\]
The important point is that $\square^{n-m-1}S$ is also a scalar, and therefore
the way the $\da\Box$ operator acts on it is exactly the same as its action on
$S$, one simply has to substitute $S\rightarrow \square^{n-m-1}S$ in
(\ref{deltaBoxS}). This is why, as we shall soon see, it becomes possible to
calculate the variation of any term of the form
\[
\int d^4x\sqrt{-g}R\Box^n R\,,
\]
that appears in (\ref{action}).

Moreover, one can use very similar methods to obtain variations such as
$\da(\Box) \Box^n R_{\mu\nu}$ and  $\da(\Box) \Box^n C_{\mu\nu\lambda\sigma}$;
since the $\Box$ operator does not affect the tensorial structure, all one needs
is to find the general form of quantities such as  $\delta(\Box)S_{\mu\nu}$
and $\delta(\Box)S_{\mu\nu\lambda\sigma}$. Also note, (\ref{deltaboxn}) actually
holds for any $S$ and $T$ irrespective of their tensorial indices. Thus, one can
adopt exactly the same procedure as for the scalar $S,T$'s to obtain the
variations involving the Ricci and Weyl tensors. Below we enumerate the somewhat
cumbersome expressions for $\delta(\Box)S_{\mu\nu}$
and $\delta(\Box)S_{\mu\nu\lambda\sigma}$. The details of the derivations can be
found in
Appendix~A.
\begin{eqnarray}
&&\delta(\square)S_{\mu\nu}=-h_{\alpha\beta}S_{\mu\nu}^{;\alpha;\beta}-(h_{
\alpha\beta})^{;\beta}S_{\mu\nu}^{;\alpha}+\frac{1}{2}g^{\alpha\beta}(h_{
\alpha\beta})^{;\sigma}S_{\mu\nu;\sigma}
\nonumber\\&&
-\frac{1}{2}\left[\square(h_{\alpha\beta})\delta_{(\mu}^{\beta}S_{\;\nu)}^{
\alpha}-(h_{\alpha\beta})^{;\tau;\alpha}\delta_{(\mu}^{\beta}S_{\tau\nu)}+(h_{
\alpha\beta})_{;(\mu}^{\;;\beta}S_{\;\nu)}^{\alpha}\right]
\nonumber\\&&
-S_{\;(\nu}^{\alpha;\beta}h_{\alpha\beta;\mu)}-\delta_{(\mu}^{\beta}S_{\;\nu)}^{
\alpha;\lambda}h_{\alpha\beta;\lambda}+\delta_{(\mu}^{\beta}S_{\tau\nu)}^{
;\alpha}h_{\alpha\beta}^{\;\;;\tau}),
\label{deltaS2}
\end{eqnarray}
\begin{eqnarray}
&&\fl\delta(\Box)S_{\mu\nu\lambda\sigma}=-h_{\alpha\beta}S_{\mu\nu\lambda\sigma}^{
;\alpha;\beta}-(h_{\alpha\beta})^{;\beta}S_{\mu\nu\lambda\sigma}^{;\alpha}+\frac
{1}{2}h{}^{;\tau}S_{\mu\nu\lambda\sigma;\tau}
\nonumber\\&&\fl
-\frac{1}{2}[g^{\alpha\tau}(h_{\alpha\beta})_{;\mu}^{;\beta}S_{
\tau\nu\lambda\sigma}+g^{\alpha\tau}(h_{\alpha\beta})_{;\nu}^{;\beta}S_{
\mu\tau\lambda\sigma}+g^{\alpha\tau}(h_{\alpha\beta})_{;\lambda}^{;\beta}S_{
\mu\nu\tau\sigma}+g^{\alpha\tau}(h_{\alpha\beta})_{;\sigma}^{;\beta}S_{
\mu\nu\lambda\tau}]
\nonumber\\&&\fl
-\left[g^{\alpha\tau}(h_{\alpha\beta})_{;\mu}S_{\tau\nu\lambda\sigma}^{;\beta}
+g^{\alpha\tau}(h_{\alpha\beta})_{;\nu}S_{\mu\tau\lambda\sigma}^{;\beta}+g^{
\alpha\tau}(h_{\alpha\beta})_{;\lambda}S_{\mu\nu\tau\sigma}^{;\beta}+g^{
\alpha\tau}(h_{\alpha\beta})_{;\sigma}S_{\mu\nu\lambda\tau}^{;\beta}\right].
\nonumber\\&&\fl
 \label{deltaS4}
 \end{eqnarray}
\subsection{Variation of the Quadratic Curvature Action}
We are now ready to compute the variation of our action (\ref{action}) and
obtain the field equations. Let us define the ``gravitational energy-momentum
tensor'' as
\[
P^{\al\bt}=-{2\over \sqrt{-g}}{\da S\over \da g_{\al\bt}}=-{2\over
\sqrt{-g}}{\da S\over h_{\al\bt}}\,.
\]
We shall compute the contribution to $P^{\al\bt}$ for the individual terms in
the action (\ref{action}) separately. We direct the reader to (\ref{P}) for the
complete field equations, should they wish to skip the technical details.
\subsubsection{Computing $\delta S_0$}
$S_0$ is nothing more than the Einstein-Hilbert action and it's variation is
well known:
\[
P_0^{\alpha\beta}=G^{\alpha\beta}=-{2\over \sqrt{-g}}{\da S_0\over
h_{\al\bt}}\,,
\]
where $G^{\alpha\beta}=R^{\alpha\beta}-\frac{1}{2}g^{\alpha\beta}R$ is the
Einstein tensor.
\subsubsection{Computing $\delta S_1$}

The next step is to compute the variation of
\[
S_{1}=\int d^{4}x\sqrt{-g}R{\cal F}_{1}(\Box)R\ .
\]
Varying this and substituting values for $\delta R$ and $\delta \sqrt{-g}$ we
find after appropriate integration by parts
\begin{eqnarray}
\delta S_1&=&\int d^{4}x\sqrt{-g}\biggl(\frac{1}{2}g^{\alpha\beta}R{\cal
F}_1R+2({\cal F}_{1}R)^{;\alpha;\beta}-2g^{\alpha\beta}\square({\cal
F}_{1}R)
\nonumber\\&&
-2R^{\alpha\beta}{\cal F}_{1}R\biggr)h_{\alpha\beta}
+\sqrt{-g}R\delta
{\cal F}_{1}R.
\end{eqnarray}
To calculate the final term, we must employ the identity in (\ref{deltaboxn}).
This gives us
\[
R\delta {\cal
F}_{1}(\Box)R=\sum_{n=1}^{\infty}\sum_{m=0}^{n-1}f_{i_{n}}R^{(m)}
\delta(\square)R^{(n-m-1)}\,,
\]
where $R^{(m)}\equiv\Box^m R$ and the analogous notation is used hereafter for
any tensor.
We then substitute our value for $\delta(\Box)$ and integrate by parts. Further
terms will cancel by noting that
\[
\label{trick}
\fl
\int d^{4}x\sqrt{-g}\sum_{n=1}^{\infty}\sum_{m=0}^{n-1}A^{(m)}B^{(n-m-1)}=\int
d^{4}x\sqrt{-g}\sum_{n=1}
^{\infty}\sum_{m=0}^{n-1}A^{(n-m-1)}B^{(m)}\,,
\]
until we arrive at the energy-momentum tensor contribution:
\begin{eqnarray}
P_{1}^{\alpha\beta}&=&4G^{\alpha\beta}{\cal F}_{1}(\Box)R+g^{\alpha\beta}R{\cal
F}_{1}(\Box)R-4\left(\nabla^{\alpha}\nabla^{\beta}-g^{\alpha\beta}\Box\right){
\cal
F}_{1}(\Box)R
\nonumber\\&&
-2\Omega_1^{\alpha\beta}+g^{\alpha\beta}(\Omega_{1\sigma}^{\;\sigma
} +\bar{\Omega}_1)\,,
\end{eqnarray}
with
\[
\Omega_{1}^{\alpha\beta}=\sum_{n=1}^{\infty}f_{1_{n}}\sum_{l=0}^{n-1}\nabla^{
\alpha}R^{(l)}\nabla^{\beta}R^{(n-l-1)},\quad\bar{\Omega}_{1}=\sum_{n=1}^{\infty
}f_{1_{n}}\sum_{l=0}^{n-1}R^{(l)}R^{(n-l)},
\]
and $P^{\alpha\beta}_1$ is defined via
\[
P^{\alpha\beta}_1=-{2\over \sqrt{-g}}{\da S_1\over h_{\al\bt}}\,.
\]

\subsubsection{Computing $\delta S_2$}
We now focus on
\[
\label{s2}
S_{2}=\int d^{4}x\sqrt{-g}\left(R^{\mu\nu}{\cal F}_{2}(\Box)R_{\mu\nu}\right)\,.
 \]

Varying the action, we find
\begin{eqnarray}
\delta S_{2}&=&\int
d^4x\sqrt{-g}\biggl[\frac{1}{2}g^{\alpha\beta}R^{\mu\nu}{\cal
F}_{2}R_{\mu\nu}-2R_{\mu}^{\beta}{\cal F}_{2}R^{\mu\alpha}+2({\cal
F}_{2}R^{\mu\beta})_{;\mu}^{\,;\alpha}
\nonumber\\&&
-\square({\cal
F}_{2}R^{\alpha\beta})
-g^{\alpha\beta}({\cal
F}_{2}R^{\mu\nu}){}_{;\mu;\nu}
-\frac{1}{2}R{\cal
F}_{2}R^{\alpha\beta}\biggr] h_{\alpha\beta}
\nonumber\\&&
+\int d^4x\sqrt{-g}R_{\mu\nu}\delta{\cal F}_{2}R^{\mu\nu}\,.
\label{varys3}
\end{eqnarray}
To compute the final term, we employ the method outlined in the previous
Subsection and in particular
(\ref{deltaS2}), here we reiterate the main steps. Using identity
(\ref{deltaboxn}) we have
\[
\sum_{n=1}^{\infty}f_{2_n}R^{\mu\nu}\delta(\square^{n})R_{\mu\nu}=\sum_{m=0}^{
n-1 }
\sum_{n=1}^{\infty}f_{2_n}R^{\mu\nu(m)}\delta(\square)R_{\mu\nu}^{(n-m-1)}\,.
\]
Then using (\ref{deltaS2}) to compute $\delta(\square)R_{\mu\nu}^{(n-m-1)}$  and
integrating by parts we find
\begin{eqnarray}
&&\fl\int d^{4}x\sqrt{-g} R^{\mu\nu}\delta{\cal
F}_{2}(\Box)S_{\mu\nu}=\int
d^{4}x\sqrt{-g}\sum_{m=0}^{n-1}\sum_{n=1}^{\infty}f_{2_{n}}\biggl[-S^{
\mu\nu(m)}S_{\mu\nu}^{(n-m-1);\alpha;\beta}
\nonumber\\&&+[S^{\mu\nu(m)}S_{\mu\nu}^{
(n-m-1);\alpha}]^{;\beta}
-\frac{1}{2}g^{\alpha\beta}[S^{\mu\nu(m)}S_{
\mu\nu;\sigma}^{(n-m-1)}]^{;\sigma}
\nonumber\\&&
-\frac{1}{2}[S^{\mu\nu(m)}S_{\;(\nu}^{\alpha(n-m-1)}]_{;\mu)}^{;\beta}
-\frac{1}{
2}\square[S^{\mu\nu(m)}\delta_{(\mu}^{\beta}S_{\;\nu)}^{\alpha(n-m-1)}]
\nonumber\\&&
+\frac{1}
{2}[S^{\mu\nu(m)}\delta_{(\mu}^{\beta}S_{\sigma\nu)}^{(n-m-1)}]^{;\sigma;\alpha}
+[S^{\mu\nu(m)}S_{(\nu}^{\alpha(n-m-1);\beta}]_{;\mu)}
\nonumber\\&&
+[S^{\mu\nu(m)}\delta_{
(\mu}^{\beta}S_{\nu)}^{\alpha(n-m-1);\sigma}]_{;\sigma}
-[S^{\mu\nu(m)}\delta_{
(\mu}^{\beta}S_{\sigma\nu}^{(n-m-1);\alpha}]^{;\sigma} \biggr]h_{\alpha\beta}\,.
\end{eqnarray}

Changing the summation order we can reduce significantly the clutter so that
for the Ricci tensor $R_{\mu\nu}$ we find

\[
\fl
\int d^{4}x\sqrt{-g}R^{\mu\nu}\delta{\cal
F}_{2}(\Box)R_{\mu\nu}=\int
d^{4}x\sqrt{-g}\left(\Omega_{2}^{\alpha\beta}-\frac{1}{2}g^{\alpha\beta}
(\Omega_{2\sigma}^{\;\sigma}+\bar{\Omega}_{2})+2\Delta_{2}^{\alpha\beta}
\right)h_{\alpha\beta}\,,
 \]
which combined with (\ref{varys3}), gives us the energy-momentum component
\begin{eqnarray}
&&\fl
P_{2}^{\alpha\beta}=-g^{\alpha\beta}R^{\mu\nu}{\cal
F}_{2}(\Box)R_{\mu\nu}+4R_{\mu}^{\beta}{\cal F}_{2}(\Box)R^{\mu\alpha}-4({\cal
F}_{2}(\Box)R^{\mu\beta})_{;\mu}^{\,;\alpha}+2\square({\cal
F}_{2}(\Box)R^{\alpha\beta})
\nonumber\\&&
+2g^{\alpha\beta}({\cal F}_{2}(\Box)R^{\mu\nu}){}_{;\mu;\nu}
-2\Omega_{2}^{\alpha\beta}+g^{\alpha\beta}(\Omega_{
2\sigma}^{\;\sigma}+\bar{\Omega}_{2})-4\Delta_{2}^{\alpha\beta}.
\label{P2}
\end{eqnarray}
Here
\[
\Omega_{2}^{\alpha\beta}=\sum_{n=1}^{\infty}f_{2_{n}}\sum_{l=0}^{n-1}R_{\nu}^{
\mu;\alpha(l)}R_{\mu}^{\nu;\beta(n-l-1)},\quad\bar{\Omega}_{2}=\sum_{n=1}^{
\infty}f_{2_{n}}\sum_{l=0}^{n-1}R_{\nu}^{\mu(l)}R_{\mu}^{\nu(n-l)}\,,
\]
\[
\Delta_{2}^{\alpha\beta}=\frac{1}{2}\sum_{n=1}^{\infty}f_{2_{n}}\sum_{l=0}^{n-1}
[R_{
\;\sigma}^{\nu(l)}R^{(\beta|\sigma|;\alpha)(n-l-1)}-R_{\;\sigma}^{\nu;(\alpha(l)
}R^{
\beta)\sigma(n-l-1)}]_{;\nu}\,.
 \]
\subsubsection{Computing $\delta S_3$}
Finally we focus on the terms involving the Weyl tensors:
\[
S_{3}=\int d^{4}x\sqrt{-g}\left(C^{\mu\nu\lambda\sigma}{\cal
F}_{3}(\Box)C_{\mu\nu\lambda\sigma}\right)\,.
\]
Varying the action we find
\[
\fl\delta S_{3}=\int
d^{4}x\frac{1}{2}\sqrt{-g}g^{\alpha\beta}h_{\alpha\beta}\left(C^{
\mu\nu\lambda\sigma}{\cal
F}_{3}(\Box)C_{\mu\nu\lambda\sigma}\right)+\sqrt{-g}\delta\left(C^{
\mu\nu\lambda\sigma}{\cal F}_{3}(\Box)C_{\mu\nu\lambda\sigma}\right)\,.
\]
Now we find the first term develops as follows
\[
\fl\delta\left(C^{\mu\nu\lambda\sigma}{\cal
F}_{3}C_{\mu\nu\lambda\sigma}\right)=\delta C^{\mu\nu\lambda\sigma}{\cal
F}_{3}C_{\mu\nu\lambda\sigma}+C^{\mu\nu\lambda\sigma}{\cal F}_{3}\delta
C_{\mu\nu\lambda\sigma}+C^{\mu\nu\lambda\sigma}\delta{\cal
F}_{3}C_{\mu\nu\lambda\sigma}\,,
\]
\[
=-4C_{\;\rho\theta\psi}^{\alpha}{\cal
F}_{3}C^{\beta\rho\theta\psi}h_{\alpha\beta}+2C^{\mu\nu\lambda\sigma}{\cal
F}_{3}\delta C_{\mu\nu\lambda\sigma}+C^{\mu\nu\lambda\sigma}\delta{\cal
F}_{3}C_{\mu\nu\lambda\sigma}\,.
\]
Next, using the definition of the Weyl tensor (\ref{weyl}), we note that
\[
2\delta C_{\;\alpha\nu\beta}^{\mu}{\cal
F}_{3}C_{\mu}^{\;\alpha\nu\beta}=2[\delta
R_{\;\alpha\nu\beta}^{\mu}-\frac{1}{2}(R_{\nu}^{\mu}h_{\alpha\beta}-R_{\beta}^{
\mu}h_{\alpha\nu})]{\cal F}_{3}C_{\mu}^{\;\alpha\nu\beta}
\]
\[
=2[\delta R_{\;\alpha\nu\beta}^{\mu}-R_{\nu}^{\mu}h_{\alpha\beta}]{\cal
F}_{3}C_{\mu}^{\;\alpha\nu\beta}\,,
\]
paying careful attention when lowering the indices of the varied Weyl tensor.
From Appendix~A we have the variation of the Riemann tensor
\[
\delta
R_{\;\alpha\nu\beta}^{\mu}=\frac{1}{2}[\delta_{\nu}^{\sigma}\delta_{\alpha}^{
\tau}(h_{\sigma\tau})_{;\beta}^{\;;\mu}-\delta_{\nu}^{\sigma}g^{\mu\tau}(h_{
\sigma\tau})_{;\beta;\alpha}+g^{\mu\sigma}\delta_{\beta}^{\tau}(h_{\sigma\tau})_
{;\alpha;\nu}-\delta_{\beta}^{\sigma}\delta_{\alpha}^{\tau}(h_{\sigma\tau})_{
\;;\nu}^{;\mu}]\,,
\]
and we know that for the Weyl tensor $C_{\;\nu\mu\lambda}^{\mu}=0$. We then find
that
\[
2C_{\mu\nu\lambda\sigma}\delta{\cal
F}_{3}C^{\mu\nu\lambda\sigma}=2\left(2R_{\mu\nu}{\cal
F}_{3}C_{\mu\nu\lambda\sigma}+({\cal
F}_{3}C_{\mu\nu\lambda\sigma})_{;\mu;\nu}\right)h_{\alpha\beta}\,.
\]
So that
\begin{eqnarray}
\delta S_{3}&=&\int
d^{4}x\sqrt{-g}\biggl[\frac{1}{2}g^{\alpha\beta}C^{\mu\nu\lambda\sigma}{\cal
F}_{3}C_{\mu\nu\lambda\sigma}-2C_{\;\rho\theta\psi}^{\alpha}{\cal
F}_{3}C^{\beta\rho\theta\psi}
\nonumber\\&&+2[2R_{\mu\nu}{\cal
F}_{3}C_{\mu\nu\lambda\sigma}+({\cal
F}_{3}C_{\mu\nu\lambda\sigma})_{;\mu;\nu}]\biggr]h_{\alpha\beta}+C^{
\mu\nu\lambda\sigma}\delta{\cal F}_{3}C_{\mu\nu\lambda\sigma}\,.
\end{eqnarray}

To find the last term we continue as in the Ricci scalar case, only now using
(\ref{deltaS2}) instead of (\ref{deltaS4}).
We integrate by parts, repeatedly and simplify using (\ref{trick}) as well as
the identity $C_{\,\mu\lambda\nu}^{\lambda}=0$ so that the final term reduces to
follows
\[
\fl
\int
d^{4}x\sqrt{-g} C^{\mu\nu\lambda\sigma}\delta{\cal
F}_{3}C_{\mu\nu\lambda\sigma}=\int
d^{4}x\sqrt{-g}\left(\Omega_{3}^{\alpha\beta}-\frac{1}{2}g^{
\alpha\beta}(\Omega_{3\gamma}^{\;\gamma}+\bar{\Omega}_{3})+4\Delta_{3}^{
\alpha\beta}\right)h_{\alpha\beta}\,,
\]
and we find the energy-momentum tensor to be
\begin{eqnarray}
&&\fl P_{3}^{\alpha\beta}
=-g^{\alpha\beta}C^{\mu\nu\lambda\sigma}{\cal
F}_{3}(\Box)C_{\mu\nu\lambda\sigma}+4C_{\;\mu\nu\sigma}^{\alpha}{\cal
F}_{3}(\square)C^{\beta\mu\nu\sigma}-4\left(2\nabla_{\mu}\nabla_{\nu}+R_{\mu\nu}
\right){\cal F}_{3}(\square)C^{\beta\mu\nu\alpha}
\nonumber\\
 &&
-2\Omega_{3}^{\alpha\beta}+g^{\alpha\beta}(\Omega_{3\gamma}^{\;\gamma}+\bar{
\Omega}_{3})-8\Delta_{3}^{\alpha\beta}.
\end{eqnarray}
Here
\[
\fl \Omega_{3}^{\alpha\beta}=\sum_{n=1}^{\infty}f_{3_{n}}\sum_{l=0}^{n-1}C_{
\;\nu\lambda\sigma}^{\mu;\alpha(l)}C_{\mu}^{\;\nu\lambda\sigma;\beta(n-l-1)},
\quad\bar{\Omega}_{3}=\sum_{n=1}^{\infty}f_{3_{n}}\sum_{l=0}^{n-1}C_{
\;\nu\lambda\sigma}^{\mu(l)}C_{\mu}^{\;\nu\lambda\sigma(n-l)}\,,
\]
\[
\fl \Delta_{3}^{\alpha\beta}=\frac{1}{2}\sum_{n=1}^{\infty}f_{3_{n}}\sum_{l=0}^{n-1}
[C_{
\;\;\;\sigma\mu}^{\lambda\nu(l)}C_{\lambda}^{\;(\beta|\sigma\mu|;\alpha)(n-l-1)}
-C_{
\;\;\;\sigma\mu}^{\lambda\nu\;\;;(\alpha(l)}C_{\lambda}^{
\;\beta)\sigma\mu(n-l-1)}
]_{;\nu}\,.
\]
\subsection{The Complete Field Equations}
Following from this we find the equation of motion for the full action $S$ in
(\ref{action}) to be a combination of $S_0,\;S_1,\;S_2$ and $S_3$ above
\begin{eqnarray}
&&\fl P^{\alpha\beta}=G^{\alpha\beta}+4G^{\alpha\beta}{\cal
F}_{1}(\Box)R+g^{\alpha\beta}R{\cal
F}_1(\Box)R-4\left(\triangledown^{\alpha}\nabla^{\beta}-g^{\alpha\beta}
\square\right){\cal F}_{1}(\Box)R
\nonumber\\&&
-2\Omega_{1}^{\alpha\beta}+g^{\alpha\beta}(\Omega_{1\sigma}^{\;\sigma}+\bar{
\Omega}_{1}) +4R_{\mu}^{\alpha}{\cal F}_2(\Box)R^{\mu\beta}
\nonumber\\&&
-g^{\alpha\beta}R_{\nu}^{\mu}{\cal
F}_{2}(\Box)R_{\mu}^{\nu}-4\triangledown_{\mu}\triangledown^{\beta}({\cal
F}_{2}(\Box)R^{\mu\alpha})
+2\square({\cal
F}_{2}(\Box)R^{\alpha\beta})
\nonumber\\&&
+2g^{\alpha\beta}\triangledown_{\mu}\triangledown_{
\nu}({\cal F}_{2}(\Box)R^{\mu\nu})
-2\Omega_{2}^{\alpha\beta}+g^{\alpha\beta}(\Omega_{2\sigma}^{\;\sigma}+\bar{
\Omega}_{2}) -4\Delta_{2}^{\alpha\beta}
\nonumber\\&&
-g^{\alpha\beta}C^{\mu\nu\lambda\sigma}{\cal
F}_{3}(\Box)C_{\mu\nu\lambda\sigma}+4C_{\;\mu\nu\sigma}^{\alpha}{\cal {\cal
F}}_{3}(\square)C^{\beta\mu\nu\sigma}
\nonumber\\&&
-4(R_{\mu\nu}+2\triangledown_{\mu}
\triangledown_{\nu})({\cal {\cal F}}_{3}(\square)C^{\beta\mu\nu\alpha})
-2\Omega_{3}^{\alpha\beta}+g^{\alpha\beta}(\Omega_{3\gamma}^{\;\gamma}+\bar{
\Omega}_{3}) -8\Delta_{3}^{\alpha\beta}
\nonumber\\&&
=T^{\al\bt}\,,
\label{P}
\end{eqnarray}
where $T^{\al\bt}$ is the stress energy tensor for the matter components in the
universe and we have defined the following symmetric tensors:
\[
\fl\Omega_{1}^{\alpha\beta}=\sum_{n=1}^{\infty}f_{1_{n}}\sum_{l=0}^{n-1}\nabla^{
\alpha}R^{(l)}\nabla^{\beta}R^{(n-l-1)},\quad\bar{\Omega}_{1}=\sum_{n=1}^{\infty
}f_{1_{n}}\sum_{l=0}^{n-1}R^{(l)}R^{(n-l)},
\]
\[
\fl\Omega_{2}^{\alpha\beta}=\sum_{n=1}^{\infty}f_{2_{n}}\sum_{l=0}^{n-1}R_{\nu}^{
\mu;\alpha(l)}R_{\mu}^{\nu;\beta(n-l-1)},\quad\bar{\Omega}_{2}=\sum_{n=1}^{
\infty}f_{2_{n}}\sum_{l=0}^{n-1}R_{\nu}^{\mu(l)}R_{\mu}^{\nu(n-l)}\,,
\]
\[
\fl\Delta_{2}^{\alpha\beta}=\frac{1}{2}\sum_{n=1}^{\infty}f_{2_{n}}\sum_{l=0}^{n-1}
[R_{
\;\sigma}^{\nu(l)}R^{(\beta|\sigma|;\alpha)(n-l-1)}-R_{\;\sigma}^{\nu;(\alpha(l)
}R^{
\beta)\sigma(n-l-1)}]_{;\nu}\,,
 \]
\[
\fl\Omega_{3}^{\alpha\beta}=\sum_{n=1}^{\infty}f_{3_{n}}\sum_{l=0}^{n-1}C_{
\;\nu\lambda\sigma}^{\mu;\alpha(l)}C_{\mu}^{\;\nu\lambda\sigma;\beta(n-l-1)},
\;\bar{\Omega}_{3}=\sum_{n=1}^{\infty}f_{3_{n}}\sum_{l=0}^{n-1}C_{
\;\nu\lambda\sigma}^{\mu(l)}C_{\mu}^{\;\nu\lambda\sigma(n-l)}\,,
\]
\[
\fl\Delta_{3}^{\alpha\beta}=\frac{1}{2}\sum_{n=1}^{\infty}f_{3_{n}}\sum_{l=0}^{n-1}
[C_{
\;\;\;\sigma\mu}^{\lambda\nu(l)}C_{\lambda}^{\;(\beta|\sigma\mu|;\alpha)(n-l-1)}
-C_{
\;\;\;\sigma\mu}^{\lambda\nu\;\;;(\alpha(l)}C_{\lambda}^{
\;\beta)\sigma\mu(n-l-1)}
]_{;\nu}\,.
\]
The trace equation is often particularly useful and below we provide it for the
general action (\ref{action}):
\begin{eqnarray}
&&\fl P=-R+12\square{\cal F}_{1}(\Box)R+2\square({\cal
F}_{2}(\Box)R)+4\triangledown_{\mu}\triangledown_{\nu}({\cal
F}_{2}(\Box)R^{\mu\nu})
\nonumber\\ &&
+2(\Omega_{1\sigma}^{\;\sigma}+2\bar{\Omega}_{1})+2(\Omega_{2\sigma}^{\;\sigma}
+2\bar{\Omega}_{2})+2(\Omega_{3\sigma}^{\;\sigma}+2\bar{\Omega}_{3}
)-4\Delta_{2\sigma}^{\;\sigma}-8\Delta_{3\sigma}^{\;\sigma}
\nonumber\\&&
=T\equiv
g_{\al\bt}T^{\al\bt}\,.
\label{trace}
\end{eqnarray}
It is worth noting that we have checked special cases of our result
against previous
work in sixth order gravity given in \cite{Decanini} and found them to be
equivalent at least to the cubic order (see Appendix~C for details).

\section{Checks and Comparisons}
The mere complexity of the derived field equations for the Generalized Quadratic
Curvature gravity warrants that we perform consistency tests and comparisons
with previous known results. This is going to be the focus of this Section.  We
will start with testing Bianchi identities.
\subsection*{3.1 Testing Bianchi Identities}
The stress energy tensor for any covariant gravitational action must satisfy
what are called the Bianchi identities:
\[
P^{\alpha\beta}_{;\beta}=0\,.
\]
Therefore, a strong test of the validity of the equation of motion (52)
would be to take the covariant derivative and to check explicitly if it
vanishes. Similar checks were also performed in [25] when
looking at actions involving only the scalar curvature. In fact, it should be
noted that the covariant derivative of each of
$P^{\alpha\beta}_0,P^{\alpha\beta}_1,P^{\alpha\beta}_2,P^{\alpha\beta}_3$ should
vanish individually as they are independent of each other. Clearly
$P^{\;\alpha\beta}_{0;\beta}=0$ as $G^{\alpha\beta}_{;\beta}\equiv0$. Now, let
us check whether the other expressions satisfy these conditions. We will focus on $P_1^{\alpha\beta}$. We have from (52):
\begin{eqnarray}
&&\fl P_{1}^{\alpha\beta}=-4G^{\alpha\beta}{\cal F}_{1}(\Box)R+g^{\alpha\beta}R{\cal
F}_{1}(\Box)R-4\left(\nabla^{\alpha}\nabla^{\beta}-g^{\alpha\beta}
\square\right){\cal F}_{1}(\Box)R
\nonumber\\&&\fl
+\sum_{n=1}^{\infty}f_{1_{n}}\sum_{l=0}^{n-1}\left\{
-2\nabla^{\alpha}R^{(l)}\nabla^{\beta}R^{(n-l-1)}
+g^{\alpha\beta}\nabla^{\sigma}R^{(l)}\nabla_{\sigma}R^{(n-l-1)}+g^{\alpha\beta}
R^{(l)}R^{(n-l)}\right\}\,.
\end{eqnarray}

Next, we take the covariant derivative and cancel like terms
\begin{eqnarray}
&&\fl P_{1;\beta}^{\alpha\beta} =4R_{\;\sigma}^{\alpha}\nabla^{\sigma}{\cal F}_{1}(\Box)R-4\nabla^{\sigma}\nabla^{\alpha}\nabla_{\sigma}{\cal F}_{1}(\Box)R+4\nabla^{\alpha}\square{\cal F}_{1}(\Box)R\\&&\fl
 +\sum_{n=1}^{\infty}f_{1_{n}}\sum_{l=0}^{n-1}\biggl[\nabla_{\sigma}R^{(l)}\nabla^{\alpha}\nabla^{\sigma}R^{(n-l-1)}-\nabla^{\sigma}\nabla^{\alpha}R^{(l)}\nabla_{\sigma}R^{(n-l-1)}
 \\&&\fl +R^{(l)}\nabla^{\alpha}R^{(n-l)}-\nabla^{\alpha}R^{(l)}R^{(n-l)}\biggr]
\end{eqnarray}
Before using the following identity which comes from the general definition of a covariant derivative acting on a tensor
\[
\label{comm}
[\nabla_a,\nabla_b]\lambda^c = R^c_{\;dab}\lambda^d
\]
which gives us
\[
\nabla^{\sigma}\nabla^{\alpha}\nabla_{\sigma}{\cal F}_{1}(\Box)R=\nabla^{\alpha}\Box{\cal F}_{1}(\Box)R+R_{\sigma}^{\;\alpha}\nabla^{\sigma}{\cal F}_{1}(\Box)R
 \]
 Substituting, we find
  \begin{eqnarray}
 &&\fl
 T_{;\beta}^{\alpha\beta} =\nabla^{\alpha}R{\cal F}_{1}(\Box)R-R\nabla^{\alpha}{\cal F}_{1}(\Box)R
 +\sum_{n=1}^{\infty}f_{1_{n}}\sum_{l=0}^{n-1}\biggl[\nabla_{\sigma}R^{(l)}\nabla^{\alpha}\nabla^{\sigma}R^{(n-l-1)}
 \\&& \fl-\nabla^{\sigma}\nabla^{\alpha}R^{(l)}\nabla_{\sigma}R^{(n-l-1)}+R^{(l)}\nabla^{\alpha}R^{(n-l)}-\nabla^{\alpha}R^{(l)}R^{(n-l)}\biggr]
\end{eqnarray}

Finally, using the following technical trick
\[
\sum_{n}\sum_{m}A^{(m)}B^{(n)}=\sum_{n}\sum_{m}A^{(n)}B^{(m)}
 \]
 along with the definition for ${\cal F}_i(\Box)=\sum^{\infty}_{n=0}f_{i_n}\Box^n$, we find that all terms cancel and so the Bianchi identity is satisfied. The same method can be applied to $S_2$ and $S_3$ by amending (\ref{comm}) so that the commutator acts upon tensors of different type with the addition of many more terms. Suffice to say, when all these terms are accounted for the Bianchi identity is indeed satisfied.
\subsection{Gravitation in a Weak-Field Limit}
Finally, we perform a last check of our equations involving linearized field
equations around the Minkowski space-time that was derived for the action
(\ref{action}) in~\cite{Biswas:2011ar}. The algorithm is to look at
fluctuations around the
Minkowski space-time:
\[
g_{\mu\nu}=\eta_{\mu\nu}+h_{\mu\nu}\,,
\label{lin-metric}
\]
$\eta^{\mu\nu}$ being the Minkowski metric. We then compute the field equations
keeping terms only up to $\cO(h)$ and compare with the equations derived
in~\cite{Biswas:2011ar}.

To accomplish this we need to compute all the relevant tensorial quantities up
to linear order in $h_{\mu\nu}$. We start with the
 inverse of the metric:
\[
g^{\mu\nu}=\eta^{\mu\nu}-h^{\mu\nu}\,.
\label{lin-inverse}
\]

Using (\ref{lin-metric}) and (\ref{lin-inverse}) we then find the Christoffel
symbols:
\[
\Gamma^{\lambda}_{\mu\nu}=\frac{1}{2}\eta^{\lambda\tau}(h_{\mu\tau,\nu}+h_{
\tau\nu,\mu}-h_{\mu\nu,\tau})
+{\cal O}(h^2)\,,
\]
where ${\cal O}(h^{2})$ represents products of $h^{\mu\nu}$ which can be ignored
in the weak-field limit. Substituting the expression for the Christoffel
symbol into the general definitions for the Riemann tensor, Ricci tensor and
curvature scalar we find the weak limit of these to be as follows:
\[
\label{Riew}
R_{\rho\mu\sigma\nu}=\frac{1}{2}\left(\partial_{\sigma}\partial_{\mu}h_{\rho\nu}
+\partial_{\nu}\partial_{\rho}h_{\mu\sigma}-\partial_{\nu}\partial_{\mu}h_{
\rho\sigma}-\partial_{\sigma}\partial_{\rho}h_{\mu\nu}\right)\,,
\]
\[
R_{\mu\nu}=\frac{1}{2}\left(\partial_{\sigma}\partial_{\mu}h_{\nu}^{\sigma}
+\partial_{\nu}\partial_{\sigma}h_{\mu}^{\sigma}-\partial_{\nu}\partial_{\mu}
h-\Box h_{\mu\nu}\right)\,,
\]
\[
\label{Rw}
R=\partial_{\mu}\partial_{\nu}h^{\mu\nu}-\square h\,,
\]
where $\square=\eta^{\mu\nu}\partial_{\mu}\partial_{\nu}$ and
$h=\eta^{\mu\nu}h_{\mu\nu}$.

From (\ref{P}), we find our equation of motion in the
weak limit up to the linear order to be
\begin{eqnarray}
P^{\alpha\beta}&=&G^{\alpha\beta}-4\left(\triangledown^{\alpha}\triangledown^{
\beta}-\eta^{\alpha\beta}\square\right){\cal
F}_{1}R-4\triangledown_{\mu}\triangledown^{\beta}({\cal
F}_{2}R^{\mu\alpha})+2\square({\cal
F}_{2}R^{\alpha\beta})
\nonumber\\&&
+2\eta^{\alpha\beta}\triangledown_{\mu}\triangledown_{\nu}
({\cal F}_{2}R^{\mu\nu})
-8\nabla_{\mu}\nabla_{\nu}{\cal F}_{3}(\Box)C^{\beta\mu\nu\alpha}\,.
\end{eqnarray}
One can then substitute equations (\ref{Riew}) to (\ref{Rw}) into the above
field equation, in order to find the weak-limit equation of motion
\begin{eqnarray}
\fl P^{\alpha\beta} & =&-\frac{1}{2}\left[1+2{\cal F}_{2}(\Box)\square+4{\cal
F}_{3}(\Box)\Box\right]\square h^{\alpha\beta}
\nonumber\\\fl
& & -\frac{1}{2}\left[-1-2{\cal F}_{2}(\Box)\Box-4{\cal
F}_{3}(\Box)\Box\right]\partial_{\sigma}(\partial^{\alpha}h^{\sigma\beta}
+\partial^{\beta}h^{\alpha\sigma})\nonumber\\\fl
& & -\frac{1}{2}\left[1-8{\cal F}_{1}(\Box)\square-2{\cal
F}_{2}(\Box)\square+\frac{4}{3}{\cal
F}_{3}(\Box)\Box\right]\left(\partial^{\beta}\partial^{\alpha}h+\eta^{
\alpha\beta}\partial_{\mu}\partial_{\nu}h^{\mu\nu}\right)
\nonumber\\\fl
& &
-\frac{1}{2}\left[-1+8{\cal F}_{1}(\Box)\Box+2{\cal
F}_{2}(\Box)\Box-\frac{4}{3}{\cal F}_{3}(\Box)\Box\right]\eta^{\alpha\beta}\Box
h\nonumber\\\fl
&
 & -\frac{1}{2}\left[8{\cal F}_{1}(\Box)\Box+4{\cal
F}_{2}(\Box)\square+\frac{8}{3}{\cal
F}_{3}(\Box)\Box\right]\Box^{-1}\triangledown^{\alpha}\triangledown^{\beta}
\partial_{\mu}\partial_{\nu}h^{\mu\nu}\,,
\end{eqnarray}
where we have used the definition of the Weyl tensor (\ref{weyl}) and
substituted equations (\ref{Riew}) to (\ref{Rw}) in order to compute the final
term. We may then rewrite this as
\begin{eqnarray}
&&\fl P^{\alpha\beta}=-\frac{1}{2}\biggl[a(\square)\square
h^{\alpha\beta}+b(\square)\partial_{\sigma}(\partial^{\alpha}h^{\sigma\beta}
+\partial^{\beta}h^{\sigma\alpha})
+c(\Box)\left(\partial^{\alpha}\partial^{\beta}h+\eta^{\alpha\beta}\partial_{\mu
}\partial_{\nu}h^{\mu\nu}\right)
\nonumber\\&&
+d(\Box)\eta^{\alpha\beta}\square
h+f(\Box)\square^{-1}\partial^{\alpha}\partial^{\beta}\partial_{\mu}\partial_{
\nu}h^{\mu\nu}\biggr]\,,
\end{eqnarray}
where we have defined the functions $a,b,c,d,f$ according to
Ref.~\cite{Biswas:2011ar,Biswas:2013ds}
\[
\label{a}
a(\Box)=\left[1+2{\cal F}_{2}(\Box)\square+4{\cal F}_{3}(\Box)\Box\right]\,,
\]
\[
b(\Box)=\left[-1-2{\cal F}_{2}(\Box)\Box-4{\cal F}_{3}(\Box)\Box\right]\,,
\]
\[
c(\Box)=\left[1-8{\cal F}_{1}(\Box)\square-2{\cal
F}_{2}(\Box)\square+\frac{4}{3}{\cal F}_{3}(\Box)\Box\right]\,,
\]
\[
d(\Box)=\left[-1+8{\cal F}_{1}(\Box)\Box+2{\cal
F}_{2}(\Box)\Box-\frac{4}{3}{\cal F}_{3}(\Box)\Box\right]\,,
\]
\[
\label{f}
f(\Box)=2\left[4{\cal F}_{1}(\Box)\Box+2{\cal
F}_{2}(\Box)\square+\frac{4}{3}{\cal F}_{3}(\Box)\Box\right]\,,
\]
 and we retrieve the following constraints~\cite{Biswas:2011ar,Biswas:2013ds}
\[
\label{ab}
a+b=0\,,
\]
\[
c+d=0\,,
\]
\[
\label{bcf}
b+c+f=0\,.
\]

Let us make a few comments about the comparison: First, we point out the slight
difference in the functions (\ref{a}-\ref{f}) as compared to
~\cite{Biswas:2011ar,Biswas:2013ds}. This is because we are using the ``mostly
positive'' convention for the metric as opposed to ``mostly negative'' used in
~\cite{Biswas:2011ar,Biswas:2013ds}. Secondly, we have set $M_p=1$ as opposed to
$M_p=2$ in~\cite{Biswas:2011ar,Biswas:2013ds}. Thirdly,
in~\cite{Biswas:2011ar,Biswas:2013ds} the action and the $\cF$'s were defined
using the Riemann tensor instead of the Weyl tensor that we are using here. We
have also defined the functions (\ref{a}-\ref{f}) in such a way so that the GR
limit tends to 1 (i.e. $a(\Box)\rightarrow 1$ when ${\cal F}_i(\Box)=0$).
Needless to say, once all these ``convention'' related differences are taken
into account, (\ref{a}-\ref{f}) become consistent with the functions derived in
~\cite{Biswas:2011ar,Biswas:2013ds}. This concludes the consistency tests on our
derived field equations.

\section{Ghost \& Asympototically free Theories, a special subclass}

The linearized field equations derived in the last Subsection provides us with
an insight about the quantum consistency and UV properties of such generalized
gravitational theories. Essentially the functions  $a(\Box)$ through $f(\Box)$
are related to the inverse propagators of the various metric degrees of freedom.
Since poles in propagators provides us with physical degrees of freedom, it is
possible to place simple criteria on these functions to ensure the theory
doesn't contain unwanted degrees of freedom. For instance, if one does not want
to introduce any new degrees of freedom apart from the massless graviton, then
$f$ must vanish, which implies that one only has a single undetermined function:
$a(\Box)$:
\[
a(\Box) = c(\Box) =-b(\Box)=-d(\Box) \Ra
6\cF_1(\Box)+3\cF_2(\Box)+2\cF_3(\Box)=0\,.
\]
While several different $\cF$'s can satisfy the above relation, for the purpose
of illustration let us consider the case when
\[
 \cF_3=0 \Ra {\cal
F}_1(\Box)=-\frac{1}{2}{\cal F}_2(\Box) \Ra {\cal F}_2(\Box)={a(\Box)-1\over
2\Box}\,.
\]
Further, the theory is ghost-free if $a(\Box)$ is an entire function without any
zeroes in the complex plane. Thus we obtain a special subclass of ghost-free
quadratic curvature theories for the massless graviton given by
\begin{equation}
S = \int d^4x\ \sqrt{-g}\left[{R\over 2}+R \LT{a(\Box)-1\over \Box}\RT R-2
R_{\mu\nu}
\LT{a(\Box)-1\over \Box}\RT R^{\mu\nu} \right ]\,,
\label{exponential}
\end{equation}
with $a(0)=1$ ensuring that we recover GR in the low energy Newtonian limit.

A particularly simple class which mimics the stringy gaussian non-localities is
given by
\[
\label{choice-a}
a(\Box)=e^{-{\Box\over M^2}}\,.
\]
By construction the above action contains only the graviton as physical degrees
of freedom as in GR, but contains an exponentially damped propagator in the UV
which, as was argued in~\cite{Biswas:2011ar}, can have profound consequences for
the gravitational singularities. For instance, one finds that if one considers
the static weak field limit and try to derive the Newtonian potentials
$\Phi,\Psi$ defined via
\[
\label{metric}
ds^2=-(1+2\Phi)dt^2+(1-2\Psi){d\vec r}\,{}^2\,,
\]
then one finds that they do not diverge for a large class of entire functions,
$a(\Box)$. For  $a(\Box)= e^{-\Box/M^2}$, $\Phi=\Psi\rightarrow $ to a constant
\cite{Nesseris:2009jf} as $r\rightarrow 0$, {\it i.e.} the theory is
asymptotically free.

One can read off the field equations for such theories from our expression
(\ref{P}) as
\begin{eqnarray}
&&\fl P^{\alpha\beta}=G^{\alpha\beta}+4G^{\alpha\beta}\left[\frac{e^{-\Box/M^{2}}-1}{\Box}\right]
R+g^ {\alpha\beta}R\left[\frac{e^{-\Box/M^{2}}-1}{\Box}\right]R
\nonumber\\
& &
-4\left(\triangledown^{\alpha}\nabla^{\beta}-g^{\alpha\beta}\square\right)\left[
\frac{e^{-\Box/M^{2}}-1}{\Box}\right]R-8R_{\mu}^{\alpha}\left[\frac{e^{-\Box/M^{
2}}-1}{\Box}\right]R^{\mu\beta}\nonumber\\
& &
+2g^{\alpha\beta}R_{\nu}^{\mu}\left[\frac{e^{-\Box/M^{2}}-1}{\Box}\right]R_{\mu}
^{\nu}+8\triangledown_{\mu}\triangledown^{\beta}\left(\left[\frac{e^{-\Box/M^{2}
}-1}{\Box}\right]R^{\mu\alpha}\right)\nonumber\\
& &
-4\square\left(\left[\frac{e^{-\Box/M^{2}}-1}{\Box}\right]R^{\alpha\beta}
\right)-4g^{\alpha\beta}\triangledown_{\mu}\triangledown_{\nu}\left(\left[\frac{
e^{-\Box/M^{2}}-1}{\Box}\right]R^{\mu\nu}\right)\nonumber\\
& &
+\sum_{n=1}^{\infty}\sum_{p=0}^{\infty}(-1)^{p+1}\frac{1}{(p+1)!M^{2(p+1)}}\sum_
{l=0}^{n-1}\biggl[-2\nabla^{\alpha}R^{(l)}\nabla^{\beta}R^{(n-l-1)}
\nonumber\\
& &+4R^{
\mu\nu;\alpha(l)}R_{\mu\nu}^{;\beta(n-l-1)}
+8[R_{\;\sigma}^{\nu(l)}R^{\beta\sigma;\alpha(n-l-1)}-R_{\;\sigma}^{
\nu;\alpha(l)}R^{\beta\sigma(n-l-1)}]_{;\nu}\}\nonumber\\
& &
+g^{\alpha\beta}\biggl(\nabla^{\sigma}R^{(l)}\nabla_{\sigma}R^{(n-l-1)}+R^{(l)}R^
{(n-l)}-2R^{\mu\nu;\sigma(l)}R_{\mu\nu;\sigma}^{(n-l-1)}
\nonumber\\&&
-2R_{\nu}^{\mu(l)}R_{\mu
}^{\nu(n-l)}\biggr)\biggr]\,,
\end{eqnarray}
where we have used (\ref{deltaboxn}) and (\ref{F_i}) to find
\[
f_{1_{n}}=\sum_{n=0}^{\infty}(-1)^{n+1}\frac{1}{(n+1)!M^{2(n+1)}}\,.
\]

\section{Conclusion}
To summarize, we have studied the classical equations of motion of the most
general extension of Einstein's gravity including terms which are quadratic in
curvatures. In particular these can contain an infinite set of higher derivative
terms because we allow any number of covariant derivatives in our action. Our
main result is the derivation of the field equations for these generalized
theories. We have tested our results using the Bianchi identity and compared our
results with known field equations for special subclasses of our starting action
~(\ref{action}) as well as  the weak-field limit that was derived
in~\cite{Biswas:2011ar}. We then illustrated our results by  specializing to a
subclass of non-local ghost and asymptotically free theories of gravity.

What are the motivations behind looking into such non-local extensions of
gravity? As mentioned in the introduction,  GR suffers from a problem with
infinities in the UV. For instance, classical singulatities appear in
cosmological and black hole solutions. Now, it is known that higher derivative
theories have better behaviour in the UV, but what conditions, must be set in
order to formulate an alternative theory of gravity? We know from Lovelock's
Theorem~\cite{lovelock} that in order to preserve general covariance, we must
accept higher than second derivatives of the metric~\footnote{More technically,
the theorem is true if we want a theory of gravity without giving up the metric
tensor (graviton), four-dimensional space or (0,2)-tensor symmetry.}. In order
to uphold unitarity, however, we must limit ourselves to theories of gravity
which are free from ghosts. The action ~(\ref{exponential}), for instance,
describes a theory which is just that, namely, it is non-local, contains an
infinite number of higher
derivatives which help to avoid the problem of ghosts and is asymptotically free
in the UV \cite{Biswas:2011ar}. Needless to say the theory recovers the
Newtonian gravitational potential in the IR. However, these statements have only
been robustly  verified at the linearized level around the Minkowski space-time.
The present work lays the groundwork for more revealing work around more general
space-times and possibly incorporating nonlinear gravitational effects.

For example, one may then examine the results in the light of an expanding
universe~\cite{Koshelev:2013lfm}, such as in de Sitter space. Another step
could be to derive the geodesic deviation equation and the Raychaudhuri equation
for our theory of gravity in order to find a black hole solution that is free
from singularities. As an aside, we remark that our formalisms and results may
also find applications in  gravitational research in other current areas, such
as studies involving inflation theory or alternatives to dark energy.

\ack
Special thanks to Spyridon
Talaganis for his helpful suggestions.
A.C. is funded by STFC grant no ST/K50208X/1;
A.K. is supported by an ``FWO-Vlaanderen''
postdoctoral fellowship and also
supported in part by Belgian Federal Science
Policy Office through the Interuniversity Attraction Pole P7/37, the
``FWO-Vlaanderen'' through the project G.0114.10N and the RFBR grant
11-01-00894; AM is supported by the Lancaster-Manchester-Sheffield Consortium
for Fundamental Physics under STFC grant ST/J000418/1.

\appendix
\section{Variation of the action}
\subsection{Background}
We have from the definitions of the Riemann and Ricci tensor
\[
\delta
R_{\;\mu\sigma\nu}^{\lambda}=(\delta\Gamma_{\mu\nu}^{\lambda})_{;\sigma}
-(\delta\Gamma_{\mu\sigma}^{\lambda})_{;\nu}
\]
\[
 \delta
R_{\mu\nu}=\triangledown_{\lambda}\delta\Gamma_{\mu\nu}^{\lambda}-\triangledown_
{\nu}\delta\Gamma_{\mu\lambda}^{\lambda}
\]
\[
\delta\Gamma{}_{\mu\nu}^{\lambda}=\frac{1}{2}(h_{\;\nu;\mu}^{\lambda}+h_{
\;\mu;\nu}^{\lambda}-h_{\mu\nu}^{\;\;;\lambda})
\]
Expanding, we find
\[
\delta
R_{\;\mu\sigma\nu}^{\lambda}=\frac{1}{2}(h_{\;\nu;\mu;\sigma}^{\lambda}-h_{
\mu\nu\;;\sigma}^{\;\;;\lambda}-h_{\;\sigma;\mu;\nu}^{\lambda}+h_{
\mu\sigma\;;\nu}^{\;\;;\lambda})
\]
\[
\delta
R_{\mu\nu}=\frac{1}{2}(h_{\;\nu;\mu;\lambda}^{\lambda}+h_{\mu\lambda\;;\nu}^{
\;\;;\lambda}-\square h_{\mu\nu}-h_{;\mu;\nu})
\]
For simplicity later on we put these all in terms of the metric variation
$h_{\alpha\beta}$
\[
\fl \delta
R_{\mu\nu\lambda\sigma}=\frac{1}{2}[\delta_{\lambda}^{\alpha}\delta_{\nu}^{\beta
}(h_{\alpha\beta})_{;\sigma;\mu}-\delta_{\lambda}^{\alpha}\delta_{\mu}^{\beta}
(h_{\alpha\beta})_{;\sigma;\nu}+\delta_{\mu}^{\alpha}\delta_{\sigma}^{\beta}(h_{
\alpha\beta})_{;\nu;\lambda}-\delta_{\sigma}^{\alpha}\delta_{\nu}^{\beta}(h_{
\alpha\beta})_{;\mu;\lambda}]
\]
\[
\fl \delta
R_{\mu\nu}=\frac{1}{2}[\delta_{\nu}^{\beta}(h_{\alpha\beta})_{;\mu}^{;\alpha}
+\delta_{\mu}^{\beta}(h_{\alpha\beta})_{;\nu}^{;\alpha}-\delta_{\mu}^{\alpha}
\delta_{\nu}^{\beta}\square(h_{\alpha\beta})-g^{\alpha\beta}(h_{\alpha\beta})_{
;\mu;\nu}]
\]
Then we can find $\delta R$
\[
\delta R=\delta(g^{\mu\nu}R_{\mu\nu})
=\delta g^{\mu\nu}R_{\mu\nu}+g^{\mu\nu}\delta R_{\mu\nu}
=-h_{\alpha\beta}R^{\alpha\beta}+g^{\mu\nu}\delta R_{\mu\nu}
\]
\[
\delta
R=-h_{\alpha\beta}R^{\alpha\beta}+(h_{\alpha\beta})^{;\alpha;\beta}-g^{
\alpha\beta}\square(h_{\alpha\beta})
\]
where we have used the following notations
\[
h_{\mu\nu}=-h^{\alpha\beta}g_{\alpha\mu}g_{\beta\nu}
,~h=g^{\mu\nu}h_{\mu\nu}
,~h_{\mu\nu}=\delta g_{\mu\nu}
,~\nabla_{\mu}S=S_{;\mu}
\]
In summary, for any tensor $S$ we have
\[
\fl\delta
S_{\mu\nu\lambda\sigma}=\frac{1}{2}[\delta_{\lambda}^{\alpha}\delta_{\nu}^{\beta
}(h_{\alpha\beta})_{;\sigma;\mu}-\delta_{\lambda}^{\alpha}\delta_{\mu}^{\beta}
(h_{\alpha\beta})_{;\sigma;\nu}+\delta_{\mu}^{\alpha}\delta_{\sigma}^{\beta}(h_{
\alpha\beta})_{;\nu;\lambda}-\delta_{\sigma}^{\alpha}\delta_{\nu}^{\beta}(h_{
\alpha\beta})_{;\mu;\lambda}]
\]
\[
\delta
S_{\mu\nu}=\frac{1}{2}[\delta_{\nu}^{\beta}(h_{\alpha\beta})_{;\mu}^{;\alpha}
+\delta_{\mu}^{\beta}(h_{\alpha\beta})_{;\nu}^{;\alpha}-\delta_{\mu}^{\alpha}
\delta_{\nu}^{\beta}\square(h_{\alpha\beta})-g^{\alpha\beta}(h_{\alpha\beta})_{
;\mu;\nu}]
\]
\[
\delta
S=-h_{\alpha\beta}R^{\alpha\beta}+(h_{\alpha\beta})^{;\alpha;\beta}-g^{
\alpha\beta}\square(h_{\alpha\beta})
\]
and the Christoffel symbol in terms of $h_{\alpha\beta}$ is
\[
\delta\Gamma{}_{\mu\nu}^{\lambda}=\frac{1}{2}(g^{\lambda\alpha}\delta_{\nu}^{
\beta}h_{\alpha\beta;\mu}+g^{\lambda\alpha}\delta_{\mu}^{\beta}h_{
\alpha\beta;\nu}-\delta_{\mu}^{\alpha}\delta_{\nu}^{\beta}h_{\alpha\beta}^{
\;\;;\lambda})
 \]
\subsection{$\delta(\Box)S$}
Recall
\[
\Box=g^{\mu\nu}\nabla_{\mu}\nabla_{\nu}
\]
Then we have
\[
\delta(\Box)S=\delta
g^{\mu\nu}S_{;\mu;\nu}+g^{\mu\nu}\delta(\nabla_{\mu})S_{;\nu}+g^{\mu\nu}[
\delta(\nabla_{\nu})S]_{;\mu}
\]
\[
=-h_{\alpha\beta}S^{;\alpha;\beta}+g^{\mu\nu}\delta(\nabla_{\mu})S_{;\nu}+g^{
\mu\nu}[\delta(\nabla_{\nu})S]_{;\mu}
\]
From the general definition of the covariant derivative of a tensor we deduce
the following
\[
\fl g^{\mu\nu}\delta(\nabla_{\mu})S_{;\nu}=-g^{\mu\nu}\delta\Gamma_{\mu\nu}^{\lambda
}S_{;\lambda}
\]
\[
\fl g^{\mu\nu}[\delta(\nabla_{\nu})S]_{;\mu}=0
\]
The last term vanishes in this case as $S$ is a scalar. This will not be true
for $\delta(\Box)S_{\mu\nu}$ and $\delta(\Box)S_{\mu\nu\lambda\sigma}$. We then
integrate by parts to find
\[
\delta(\Box)S=-h_{\alpha\beta}S^{;\alpha;\beta}+\frac{1}{2}g^{\alpha\beta}S_{
;\lambda}(h_{\alpha\beta})^{;\lambda}-S^{;\alpha}(h_{\alpha\beta})^{;\beta}
\]
with
\[
\delta\Gamma_{\mu\nu}^{\lambda}=\frac{1}{2}[g^{\alpha\lambda}\delta_{\mu}^{\beta
}(h_{\alpha\beta})_{;\nu}+g^{\alpha\lambda}\delta_{\nu}^{\beta}(h_{\alpha\beta}
)_{;\mu;}-\delta_{\mu}^{\alpha}\delta_{\nu}^{\beta}(h_{\alpha\beta})^{;\lambda}]
\]
\subsection{$\delta(\Box)S_{\mu\nu}$}
\label{sec:dS2}
\[
\delta(\square)S_{\mu\nu}=\delta
g^{\lambda\sigma}S_{\mu\nu;\lambda;\sigma}+g^{\lambda\sigma}\delta(\nabla_{
\lambda})S_{\mu\nu;\sigma}+g^{\lambda\sigma}[\delta(\nabla_{\sigma})S_{\mu\nu}]_
{;\lambda}
\]
\[
=-h_{\alpha\beta}S_{\mu\nu}^{;\alpha;\beta}+g^{\lambda\sigma}\delta(\nabla_{
\lambda})S_{\mu\nu;\sigma}+g^{\lambda\sigma}[\delta(\nabla_{\sigma})S_{\mu\nu}]_
{;\lambda}
\]
From the general definition of the covariant derivative of a tensor we have
\[\fl
g^{\lambda\sigma}\delta(\nabla_{\lambda})S_{\mu\nu;\sigma}=-\delta\Gamma_{
\lambda\mu}^{\tau}S_{\tau\nu}^{;\lambda}-\delta\Gamma_{\lambda\nu}^{\tau}S_{
\mu\tau}^{;\lambda}-g^{\lambda\sigma}\delta\Gamma_{\lambda\sigma}^{\tau}S_{
\mu\nu;\tau}
\]
\[\fl
g^{\lambda\sigma}\nabla_{\lambda}\delta(\nabla_{\sigma})S_{\mu\nu}
=-(\delta\Gamma_{\lambda\mu}^{\tau})^{;\lambda}S_{\tau\nu}-\delta\Gamma_{
\lambda\mu}^{\tau}S_{\tau\nu}^{;\lambda}-(\delta\Gamma_{\lambda\nu}^{\tau})^{
;\lambda}S_{\mu\tau}-\delta\Gamma_{\lambda\nu}^{\tau}S_{\mu\tau}^{;\lambda}
\]
So that
\[
\delta(\square)S_{\mu\nu}=-h_{\alpha\beta}S_{\mu\nu}^{;\alpha;\beta}-g^{
\lambda\sigma}\delta\Gamma_{\lambda\sigma}^{\tau}S_{\mu\nu;\tau}-(\delta\Gamma_{
\lambda(\mu}^{\tau})^{;\lambda}S_{\tau\nu)}-2\delta\Gamma_{\lambda(\mu}^{\tau}S_
{\tau\nu)}^{;\lambda}
\]
Expanding using
$\delta\Gamma_{\mu\nu}^{\lambda}=\frac{1}{2}[g^{\alpha\lambda}\delta_{\mu}^{
\beta}(h_{\alpha\beta})_{;\nu}+g^{\alpha\lambda}\delta_{\nu}^{\beta}(h_{
\alpha\beta})_{;\mu;}-\delta_{\mu}^{\alpha}\delta_{\nu}^{\beta}(h_{\alpha\beta}
)^{;\lambda}]$, we have
\[
\delta(\square)S_{\mu\nu}=-h_{\alpha\beta}S_{\mu\nu}^{;\alpha;\beta}-(h_{
\alpha\beta})^{;\beta}S_{\mu\nu}^{;\alpha}+\frac{1}{2}g^{\alpha\beta}(h_{
\alpha\beta})^{;\sigma}S_{\mu\nu;\sigma}
\]
\[
-\frac{1}{2}\left[\square(h_{\alpha\beta})\delta_{(\mu}^{\beta}S_{\;\nu)}^{
\alpha}-(h_{\alpha\beta})^{;\tau;\alpha}\delta_{(\mu}^{\beta}S_{\tau\nu)}+(h_{
\alpha\beta})_{;(\mu}^{\;;\beta}S_{\;\nu)}^{\alpha}\right]
\]
\[
-S_{\;(\nu}^{\alpha;\beta}h_{\alpha\beta;\mu)}-\delta_{(\mu}^{\beta}S_{\;\nu)}^{
\alpha;\lambda}h_{\alpha\beta;\lambda}+\delta_{(\mu}^{\beta}S_{\tau\nu)}^{
;\alpha}h_{\alpha\beta}^{\;\;;\tau})
\]
\subsection{$\delta(\Box)S_{\mu\nu\lambda\sigma}$}
From the definition of the D'Alembertian operator
$\Box=g^{\mu\nu}\nabla_{\mu}\nabla_{\nu}$, we have
\[
\delta(\Box)S_{\mu\nu\lambda\sigma}=\delta
g^{\kappa\tau}S_{\mu\nu\lambda\sigma;\kappa;\tau}+g^{\kappa\tau}\delta(\nabla_{
\kappa})S_{\mu\nu\lambda\sigma;\tau}+g^{\kappa\tau}[\delta(\nabla_{\tau})S_{
\mu\nu\lambda\sigma}]_{;\kappa}
\]
\[
=-h_{\alpha\beta}S_{\mu\nu\lambda\sigma}^{;\alpha;\beta}+g^{\kappa\tau}
\delta(\nabla_{\kappa})S_{\mu\nu\lambda\sigma;\tau}+g^{\kappa\tau}[
\delta(\nabla_{\tau})S_{\mu\nu\lambda\sigma}]_{;\kappa}
\]
and from the general definition of the covariant derivative of a tensor and
treating $S_{\mu\nu\lambda\sigma;\tau}$ as a $(0,5)-tensor$, we have
\begin{eqnarray}
&&\fl g^{\kappa\tau}\delta(\nabla_{\kappa})S_{\mu\nu\lambda\sigma;\tau}=-\delta\Gamma_
{\kappa\mu}^{\rho}S_{\rho\nu\lambda\sigma}^{;\kappa}-\delta\Gamma_{\kappa\nu}^{
\rho}S_{\mu\rho\lambda\sigma}^{;\kappa}-\delta\Gamma_{\kappa\lambda}^{\rho}S_{
\mu\nu\rho\sigma}^{;\kappa}-\delta\Gamma_{\kappa\sigma}^{\rho}S_{
\mu\nu\lambda\rho}^{;\kappa}
\nonumber\\&&
-g^{\kappa\tau}\delta\Gamma_{\kappa\tau}^{\rho}S_{
\mu\nu\lambda\sigma;\rho}
\end{eqnarray}
and
\[
\fl g^{\kappa\tau}[\delta(\nabla_{\tau})S_{\mu\nu\lambda\sigma}]_{;\kappa}=\left[
-\delta\Gamma_{\kappa\mu}^{\rho}S_{\rho\nu\lambda\sigma}-\delta\Gamma_{\kappa\nu
}^{\rho}S_{\mu\rho\lambda\sigma}-\delta\Gamma_{\kappa\lambda}^{\rho}S_{
\mu\nu\rho\sigma}-\delta\Gamma_{\kappa\sigma}^{\rho}S_{\mu\nu\lambda\rho}\right]
^{;\kappa}
\]
\begin{eqnarray}
&=&-(\delta\Gamma_{\kappa\mu}^{\rho})^{;\kappa}S_{\rho\nu\lambda\sigma}
-\delta\Gamma_{\kappa\mu}^{\rho}S_{\rho\nu\lambda\sigma}^{;\kappa}
-(\delta\Gamma_{\kappa\nu}^{\rho})^{;\kappa}S_{\mu\rho\lambda\sigma}
-\delta\Gamma_{\kappa\nu}^{\rho}S_{\mu\rho\lambda\sigma}^{;\kappa}
\nonumber\\&&
-(\delta\Gamma_{\kappa\lambda}^{\rho})^{;\kappa}S_{\mu\nu\rho\sigma}
-\delta\Gamma_{\kappa\lambda}^{\rho}S_{\mu\nu\rho\sigma}^{;\kappa}
-(\delta\Gamma_{\kappa\sigma}^{\rho})^{;\kappa}S_{\mu\nu\lambda\rho}
-\delta\Gamma_{\kappa\sigma}^{\rho}S_{\mu\nu\lambda\rho}^{;\kappa}
\end{eqnarray}
So that
\begin{eqnarray}
&&\delta(\Box)S_{\mu\nu\lambda\sigma}=-h_{\alpha\beta}S_{\mu\nu\lambda\sigma}^{
;\alpha;\beta}-g^{\kappa\tau}\delta\Gamma_{\kappa\tau}^{\rho}S_{
\mu\nu\lambda\sigma;\rho}
\nonumber\\&&
-\left[(\delta\Gamma_{\kappa\mu}^{\rho})^{;\kappa}S_{\rho\nu\lambda\sigma}
+(\delta\Gamma_{\kappa\nu}^{\rho})^{;\kappa}S_{\mu\rho\lambda\sigma}
+(\delta\Gamma_{\kappa\lambda}^{\rho})^{;\kappa}S_{\mu\nu\rho\sigma}
+(\delta\Gamma_{\kappa\sigma}^{\rho})^{;\kappa}S_{\mu\nu\lambda\rho}\right]
\nonumber\\&&
-2\left[\delta\Gamma_{\kappa\mu}^{\rho}S_{\rho\nu\lambda\sigma}^{;\kappa}
+\delta\Gamma_{\kappa\nu}^{\rho}S_{\mu\rho\lambda\sigma}^{;\kappa}+\delta\Gamma_
{\kappa\lambda}^{\rho}S_{\mu\nu\rho\sigma}^{;\kappa}+\delta\Gamma_{\kappa\sigma}
^{\rho}S_{\mu\nu\lambda\rho}^{;\kappa}\right]
\end{eqnarray}
Then, using
$\delta\Gamma_{\mu\nu}^{\lambda}=\frac{1}{2}[g^{\alpha\lambda}\delta_{\mu}^{
\beta}(h_{\alpha\beta})_{;\nu}+g^{\alpha\lambda}\delta_{\nu}^{\beta}(h_{
\alpha\beta})_{;\mu;}-\delta_{\mu}^{\alpha}\delta_{\nu}^{\beta}(h_{\alpha\beta}
)^{;\lambda}]$, and the Bianchi identities, we find
\begin{eqnarray}
&&\fl\delta(\Box)S_{\mu\nu\lambda\sigma}=-h_{\alpha\beta}S_{\mu\nu\lambda\sigma}^{
;\alpha;\beta}-(h_{\alpha\beta})^{;\beta}S_{\mu\nu\lambda\sigma}^{;\alpha}+\frac
{1}{2}h{}^{;\tau}S_{\mu\nu\lambda\sigma;\tau}
\nonumber\\&&\fl
-\frac{1}{2}[g^{\alpha\tau}(h_{\alpha\beta})_{;\mu}^{;\beta}S_{
\tau\nu\lambda\sigma}+g^{\alpha\tau}(h_{\alpha\beta})_{;\nu}^{;\beta}S_{
\mu\tau\lambda\sigma}+g^{\alpha\tau}(h_{\alpha\beta})_{;\lambda}^{;\beta}S_{
\mu\nu\tau\sigma}+g^{\alpha\tau}(h_{\alpha\beta})_{;\sigma}^{;\beta}S_{
\mu\nu\lambda\tau}]
\nonumber\\&&\fl
-\left[g^{\alpha\tau}(h_{\alpha\beta})_{;\mu}S_{\tau\nu\lambda\sigma}^{;\beta}
+g^{\alpha\tau}(h_{\alpha\beta})_{;\nu}S_{\mu\tau\lambda\sigma}^{;\beta}+g^{
\alpha\tau}(h_{\alpha\beta})_{;\lambda}S_{\mu\nu\tau\sigma}^{;\beta}+g^{
\alpha\tau}(h_{\alpha\beta})_{;\sigma}S_{\mu\nu\lambda\tau}^{;\beta}\right]
\end{eqnarray}

\section{Comparison with sixth-order Gravity}
Field equations for the most general gravitational action up to sixth order in
derivatives were derived in~\cite{Decanini}. This is the only action known to us
which contains $\Box$ operators acting on Weyl tensors, and hence it is useful
to check our results against field equations given in ~\cite{Decanini}. For our
action (\ref{action}), this would mean keeping terms only up to a single box,
{\it i.e.} $f_{In}=0$ for $n\geq 2$. More explicitly the field equations read
as follows
\begin{eqnarray}
&&P^{\alpha\beta}=G^{\alpha\beta}+4G^{\alpha\beta}(f_{10}+f_{11}\Box)R+g^{
\alpha\beta}R(f_{10}+f_{11}\Box)R
\nonumber\\&&
-4\left(\triangledown^{\alpha}\nabla^{\beta}-g^
{\alpha\beta}
\square\right)(f_{10}+f_{11}\Box)R
-2f_{11}\n^{\al}R\n^{\bt}R
\nonumber\\&&
+g^{\alpha\beta}f_{11}(\n^{\ga}R\n_{\ga}R+R\Box R) +4R_{\mu}^{\alpha}(f_{20}+f_{21}\Box)R^{\mu\beta}
-g^{\alpha\beta}R_{\nu}^{\mu}(f_{20}
\nonumber\\&&
+f_{21}\Box)R_{\mu}^{\nu}-4\triangledown_{
\mu}\triangledown^{\beta}((f_{20}+f_{21}\Box)R^{\mu\alpha})
+2\square((f_{20}+f_{21}\Box)R^{\alpha\beta})
\nonumber\\&&
+2g^{\alpha\beta}\triangledown_{\mu}\triangledown_{
\nu}((f_{20}+f_{21}\Box)R^{\mu\nu})
-2f_{21}\n^{\al}R^{\mu}_{\nu}\n^{\bt}R^{\nu}_{\mu}
\nonumber\\&&
+g^{\alpha\beta}f_{21}(\n^{\ga}R^{\mu}_{\nu}\n_{\ga}R^{\nu}_{\mu}+R^{\mu}_{\nu}\Box R^{\nu}_{\mu})
-4f_{21}(R^{\nu}_{\mu}\n^{\al}R^{\bt\mu}-R^{\bt\mu}\n^{\al}R^{\nu}_{\mu})_{;\nu}
\nonumber\\&&
-g^{\alpha\beta}C^{\mu\nu\lambda\sigma}(f_{30}+f_{31}\Box)C_{\mu\nu\lambda\sigma}
+4C_{\;\mu\nu\sigma}^{\alpha}(f_{30}+f_{31}\Box)C^{\beta\mu\nu\sigma}
\nonumber\\&&
-4(R_{\mu\nu}+2\triangledown_{\mu}
\triangledown_{\nu})((f_{30}+f_{31}\Box)C^{\beta\mu\nu\alpha})
-2f_{31}\n^{\al}C^{\mu\nu\rho\ga}\n^{\bt}C_{\mu\nu\rho\ga}
\nonumber\\&&
+g^{\alpha\beta}f_{31}(\n^{\al}C^{\mu\nu\rho\ga}\n^{\bt}C_{\mu\nu\rho\ga}+C^{
\mu\nu\rho\ga}\Box C_{\mu\nu\rho\ga})
\nonumber\\&&
-8f_{31}(C^{\ga\nu}{}_{\rho\mu}\n^{\al}C_{\ga}{}^{\bt\rho\mu}-C_{\ga}{}^{
\bt\rho\mu}\n^{\al}C^{\ga\nu}{}_{\rho\mu})_{;\nu}
\label{sixth}
\end{eqnarray}
 Unfortunately, this expression cannot be directly compared with the
terms in \cite{Decanini},
most specifically in their equation (2.23) because in \cite{Decanini} several
identities (given in their Section 5) were
used to convert quadratic curvature terms with a $\Box$ to cubic in
curvature without a $\Box$. The full matching
of the expressions therefore becomes a rather arduous task which is not
particularly illuminating and
we do not include this here. Nevertheless we have explicitly checked
that expansions up to cubic order around the Minkowski background
perfectly match.


\end{document}